\documentclass{emulateapj}
\usepackage{apjfonts}
\usepackage{graphicx}
\usepackage{amsmath}
\usepackage{multirow}
\begin{document}
\bibliographystyle{apj}
\shorttitle{STRUCTURE OF 2MASS CLUSTERS}
\shortauthors{BLACKBURNE \& KOCHANEK}
\slugcomment{Accepted to ApJ}
\title{The Structure of 2MASS Galaxy Clusters}

\author{Jeffrey~A.~Blackburne\altaffilmark{1} \& 
  Christopher~S.~Kochanek\altaffilmark{1,2}}

\altaffiltext{1}{Department of Astronomy, The Ohio State University,
  140 West 18th Avenue, Columbus, OH 43210, USA}
\altaffiltext{2}{Center for Cosmology and AstroParticle Physics, The
  Ohio State University, 191 West Woodruff Avenue, Columbus, OH 43210,
  USA}
\begin{abstract}

We use a sample of galaxies from the Two Micron All Sky Survey (2MASS)
Extended Source Catalog to refine a matched filter method of finding
galaxy clusters that takes into account each galaxy's position,
magnitude, and redshift if available. The matched filter postulates a
radial density profile, luminosity function, and line-of-sight
velocity distribution for cluster galaxies. We use this method to
search for clusters in the galaxy catalog, which is complete to an
extinction-corrected \mbox{$K$-band} magnitude of $13.25$ and has
spectroscopic redshifts for roughly 40\% of the galaxies, including
nearly all brighter than $K=11.25$. We then use a stacking analysis to
determine the average luminosity function, radial distribution, and
velocity distribution of cluster galaxies in several richness classes,
and use the results to update the parameters of the matched filter
before repeating the cluster search. We also investigate the
correlations between a cluster's richness and its velocity dispersion
and core radius, using these relations to refine priors that are
applied during the cluster search process. After the second cluster
search iteration, we repeat the stacking analysis. We find a cluster
galaxy luminosity function that fits a Schechter form, with parameters
$M_{K*}-5\log h=-23.64\pm0.04$ and $\alpha=-1.07\pm0.03$. We can
achieve a slightly better fit to our luminosity function by adding a
Gaussian component on the bright end to represent the brightest
cluster galaxy (BCG) population. The radial number density profile of
galaxies closely matches a projected Navarro-Frenk-White (NFW) profile
at intermediate radii, with deviations at small radii due to
well-known cluster centering issues and outside the virial radius due
to correlated structure. The velocity distributions are Gaussian in
shape, with velocity dispersions that correlate strongly with
richness.
\end{abstract}


\section{Introduction}
\label{sec:intro}

As the most massive structures known that are in dynamical
equilibrium, clusters of galaxies are useful for studies of
large-scale structure \citep[e.g.,][]{Bahcall:1988p631,
Einasto:2001p2222, Yang:2005p608, Papovich:2008p206}, as well as for
galaxy formation and evolution \citep[e.g.,][]{Dressler:1992p1,
Goto:2003p601} and for constraining cosmological parameters
\citep[e.g.,][]{Henry:2000p565, Allen:2008p879, Rozo:2010p645}. The
problem of finding clusters of galaxies has been attacked from several
angles. The oldest method is simply to look for overdensities in the
two-dimensional distribution of galaxies on the sky
\citep[e.g.,][]{Abell:1958p211, Zwicky:1968p3267, Abell:1989p1}. This
method faces difficulties caused by line-of-sight interlopers, a
problem which has been greatly ameliorated in recent years by
photometric (so-called 2.5-dimensional) and spectroscopic redshift
surveys. Several algorithms have been used to analyze such data,
including the percolation (or friends-of-friends) algorithm
\citep[e.g.,][]{Huchra:1982p423, Crook:2007p790}, the red sequence
method \citep[e.g.,][]{Gladders:2000p2148, Koester:2007p221} and the
matched filter method \citep[e.g.,][]{Postman:1996p615,
Kepner:1999p78, Kochanek:2003p161, Dong:2008p868,
Szabo:2010p249}. Other fruitful strategies include searching for the
thermal X-ray emission of the hot intra-cluster gas
\citep[e.g.,][]{Gioia:1990pL35, Ebeling:1998p881, Bohringer:2001p826,
Mullis:2003p154} and searching for the weak lensing signature of
clusters \citep[e.g.,][]{Schneider:1996p837, Wittman:2001pL89,
Sheldon:2009p2217}. The most recently developed approach is to search
for the thermal Sunyaev-Zeldovich decrement in the cosmic microwave
background caused by this same gas \citep[e.g.,][]{Carlstrom:2000p148,
LaRoque:2003p559, Staniszewski:2009p32}. These methods can be used to
estimate cluster masses, which are important for comparison with
theory. But optical and infrared (IR) surveys primarily measure
cluster richness, a quantity which, though correlated with cluster
mass, has significant scatter at fixed mass.

In this paper we follow up the work of \citet[][K03
  hereafter]{Kochanek:2003p161}, who use a matched filter algorithm
based largely on the earlier approach of \citet{Kepner:1999p78} to
find clusters of galaxies in the Two Micron All Sky Survey (2MASS)
Extended Source Catalog \citep{Jarrett:2000p2498,
  Skrutskie:2006p1163}. This method makes use of the expected
properties of galaxy clusters --- specifically, their shapes in
angular and redshift space, and the luminosity function of their
members. Our aim in this paper is to evaluate and update the
parameters of the matched filter in order to increase the completeness
and purity of the resulting cluster catalog, and to obtain richness
estimates that are as accurate as possible. The large number of
clusters in the 2MASS catalog enable us to use a stacking analysis to
find their average properties. We use an iterative strategy: first, we
search for clusters using a filter very similar to that of K03, then
we stack the resulting clusters to determine their average radial
density profile, velocity distribution, and luminosity function as a
function of richness. Using these properties, we then refine the
matched filter and repeat the cluster search and stacking analysis,
adopting the results of this second iteration as our best estimate of
the average properties of the cluster sample. Throughout this work we
refer to these as the first and second cluster search iterations.

\begin{figure}
\includegraphics[width=\columnwidth]{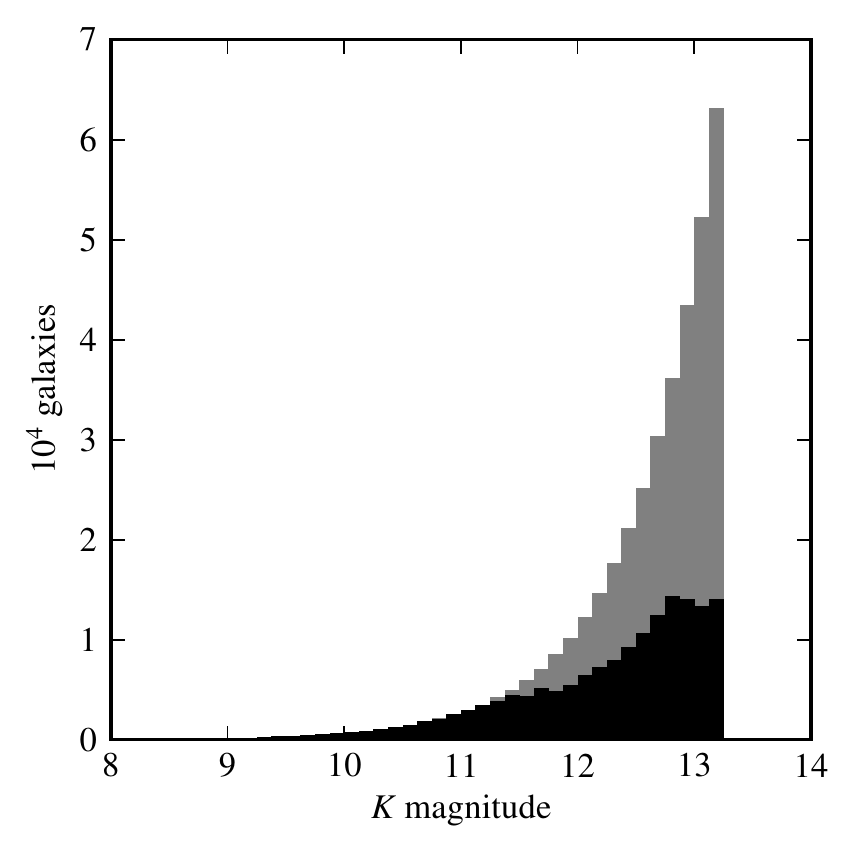}
\caption{Apparent magnitude distribution of all input galaxies (gray)
  and input galaxies with spectroscopic redshift measurements
  (black).}
\label{fig:maghist}
\end{figure}

K03 used a few modifications to their likelihood function to stabilize
parameter estimation and reflect prior knowledge; we incorporate and
expand these. In particular, one set of priors makes use of the
relationship between a cluster's richness and its core radius and
velocity dispersion. Since our cluster search method produces
estimates of these quantities for each cluster, we use the
correlations we observe in the first cluster search iteration to tune
this set of priors for the second iteration.

\begin{figure}
\includegraphics[width=\columnwidth]{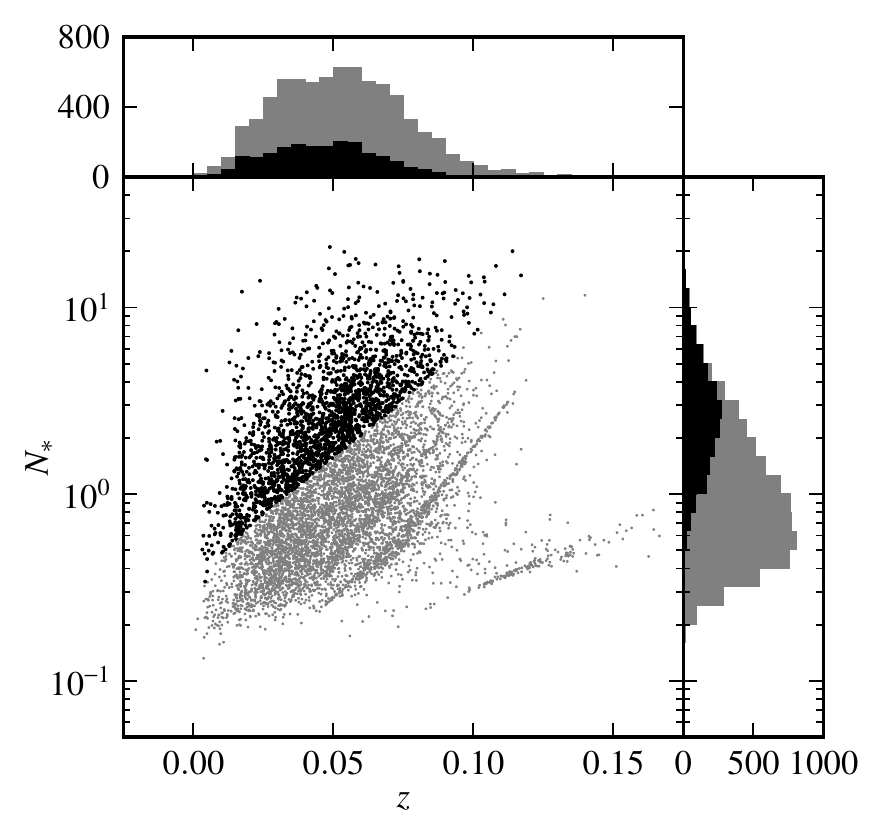}
\caption{Richness $N_*$ and redshift $z$ of the cluster candidates
  resulting from our second cluster search iteration. The black points
  are those clusters for which $N_{*666} > 3$, a sample which K03
  found to be highly pure. The black histograms correspond to these
  clusters, while the gray histograms represent all clusters with
  $\Delta \ln \mathcal{L} \ge 5$. The search is deliberately extended
  to low $\Delta \ln \mathcal{L}$ and $N_*$ to allow for later
  investigation of false positives.}
\label{fig:zrichscatter}
\end{figure}

We use a deeper version of the galaxy catalog used by K03: a
flux-limited selection from the 2MASS Extended Source Catalog with $K
< 13.25$ and Galactic latitudes $|b|>6^{\circ}$. The $20$ mag
arcsec$^{-1}$ isophotal \mbox{$K$-band} magnitudes are corrected for
Galactic extinction using the \citet{Schlegel:1998p525} extinction
model. Of our sample of 380360 galaxies, 161030 have spectroscopic
redshifts from a preliminary version of 2MASS Redshift Survey
\citep[2MRS,][]{Huchra:2011p0}. The 2MRS contains not only redshifts
measured by \citet{Huchra:2011p0}, but also compiles redshifts from
many other sources. Out of about 950, some of the most important
contributors are the Six-degree Field Galaxy Redshift Survey
\citep[e.g.,][]{Jones:2004p747}, the Sloan Digital Sky Survey
\citep[e.g.,][]{Abazajian:2005p1755}, the Two-degree Field Galaxy
Redshift Survey \citep[e.g.,][]{Colless:2001p1039}, the Center for
Astrophysics Redshift Survey \citep[e.g.,][]{Huchra:1983p89,
  Falco:1999p438}, the Las Campanas Redshift Survey
\citep[LCRS,][]{Shectman:1996p172}, the ESO Nearby Abell Cluster
Survey \citep{Katgert:1998p399}, the Southern Sky Redshift Survey
\citep{daCosta:1998p1}, and others
\citep[e.g.,][]{Bottinelli:1990p391, Loveday:1996p201}. The complete
list of sources is given by \citet{Huchra:2011p0}. Essentially all of
the 23046 galaxies with $K < 11.25$ have redshift measurements, as do
138157 galaxies fainter than that limit. The apparent magnitude
distribution of our galaxy sample is shown in
Figure~\ref{fig:maghist}. In our second cluster search iteration, we
find 7624 cluster candidates that exceed our likelihood threshold,
most with redshifts below $0.1$. The clusters are essentially
characterized by the intrinsic richness $N_*$, as well as by an
``apparent richness'' $N_g$, which declines with distance as galaxies
below the apparent magnitude limit drop out of the galaxy catalog. We
show the distribution of the cluster candidates in redshift and
richness in Figure~\ref{fig:zrichscatter}. Together with the
likelihood values of individual clusters, the completeness and purity
of the cluster catalog decline as $N_g$ decreases. They, as well as
mass estimates and comparisons to other cluster surveys, are the
subject of a forthcoming paper.

In Section~\ref{sec:model} we describe our galaxy cluster model, which
includes the spatial and velocity distribution of cluster galaxies and
their luminosity function.  We use this model as a matched filter to
find clusters. In Section~\ref{sec:stacking} we describe the stacking
technique with which we determine the average properties of the
clusters in order to update the model. In Section~\ref{sec:priors} we
describe the priors that we use the modify the likelihood function,
and use the initial sample of clusters to update some of their
parameters. Finally, in Section~\ref{sec:conclusions} we conclude. The
work described in K03 used a cosmological model with $\Omega_M = 1$
and $\Omega_\Lambda = 0$. The details of the cosmological model are
not very important for our purposes since the 2MASS galaxies are
nearby, but for this work we use a cosmology with $\Omega_M = 0.3$ and
$\Omega_\Lambda = 0.7$. We use the usual parameterization for the
Hubble constant, with $H_0 = 100h~\mathrm{km}~\mathrm{s}^{-1}~
\mathrm{Mpc}^{-1}$.

\section{The Matched Filter Method}
\label{sec:model}

Following the method of \citet{Kepner:1999p78} as expanded by K03, we
search for clusters by computing a likelihood which is the convolution
of the observed distribution of galaxies (in angular, redshift, and
luminosity space) with a filter tuned to match the ``shape'' of galaxy
clusters. This convolution smooths out the small-scale details of
individual galaxy locations but maximizes the signal due to actual
cluster-shaped galaxy overdensities. This method has several
advantages. It provides estimates for the likelihood of each detected
cluster, as well as for the membership probabilities of each nearby
galaxy. In its K03 realization, it also provides best-fit values and
uncertainties for cluster properties such as richness and velocity
dispersion, and it is flexible enough to handle galaxies with or
without redshift and color information; this is important for our
sample, where not every galaxy has a redshift measurement.

\subsection{The Cluster Model}
We add clusters to the catalog in an iterative fashion, at each step
evaluating the likelihood function at the position of each galaxy in
the sample and adding a cluster centered on the galaxy with the
highest likelihood. The likelihood function is constructed from the
probabilities of the nearby galaxies to be cluster members or
non-member field galaxies; these probabilities make up the matched
filter. For a proposed $n_c$-th galaxy cluster, the change in
likelihood is
\begin{equation}
\label{eqn:likelihood}
\Delta \ln \mathcal{L}(n_c) = -N_{*n_c} A_{n_c} + 
\sum_{i} \ln \left[
\frac{P_f(i)+\sum_{k=1}^{n_c} P_c(i,k)}
     {P_f(i)+\sum_{k=1}^{n_c-1} P_c(i,k)} \right] ~,
\end{equation}
where the sum in $i$ is over all galaxies within the sampling radius
$R_\mathrm{samp} = 1.0\,h^{-1}$ Mpc. The term $-N_{*n_c} A_{n_c}$ is
an estimate of the total number of member galaxies we expect to be
visible within $R \le R_\mathrm{samp}$ of the proposed cluster, given
the survey magnitude limit; we give an expression for it in
Equation~(\ref{eqn:Aexpect}). The definitions of $P_f$ and $P_c$ are
given in Equations~(\ref{eqn:fieldprob}) and (\ref{eqn:clusterprob}),
respectively. The expression for the likelihood in
Equation~(\ref{eqn:likelihood}) is derived in Appendix C2 of
\citet{Kepner:1999p78}. The likelihood depends on the richness,
redshift, core radius, and velocity dispersion of the candidate
cluster, mostly through $P_c$; we optimized these values using the
Markov Chain Monte Carlo (MCMC) method while evaluating the
likelihood. The clusters labeled $k=1 \cdots n_c-1$ are already in the
cluster catalog, and have fixed properties. We account for previously
found clusters via their inclusion in the denominator of the last term
in the equation; this effectively removes them from the density field
in a manner similar to the \textsc{Clean} algorithm of radio astronomy
\citep{Hogbom:1974p417}. We stop iterating when no cluster is found
that increases the likelihood by more than a predetermined cutoff
value. We choose $\Delta \ln \mathcal{L} = 5$ as our cutoff; this is
low enough that most of the low-likelihood cluster candidates are in
fact false positives (see K03). In Section~\ref{sec:stacking} we select a
relatively pure subset of clusters for analysis, all of which have
likelihoods well above the cutoff. The likelihood of a given cluster
correlates well with $N_g$, the number of member galaxies brighter than the
survey limit, and only weakly with the actual richness $N_*$.

The probability of finding a field galaxy with a given absolute $K$
magnitude $M_K$ and redshift $z$ in some infinitesimal portion of the
sky is
\begin{equation}
\label{eqn:fieldprob}
P_f = 0.4 \ln(10) D_C^2(z) \frac{dD_C}{dz} \phi_f(M_K) ~,
\end{equation}
where $D_C$ is the comoving distance and $\phi_f$ is the field galaxy
luminosity function. When the redshift of the galaxy is not known, we
average $P_f$ over the range $0 \le z < 1$; this is effectively the
differential number count of the 2MASS survey. We follow K03 in
adopting the luminosity function of \citet{Kochanek:2001p566} for
$\phi_f$. It is a Schechter function \citep{Schechter:1976p297}, with
parameters $M_{K*} = -23.39$ and $\alpha=-1.09$, and normalization
$n_* = 1.16\times10^{-2} h^3$ Mpc$^{-3}$. We also use this luminosity
function for cluster galaxies in our first search iteration; see
Equation~(\ref{eqn:clusterlf}). In the second iteration we update the
cluster luminosity function, but the field luminosity function stays
unchanged. Following the example of K03, we calculate the absolute
magnitudes using an effective distance modulus
\begin{equation}
\mathcal{D}(z) = K - M_K \equiv 5\log(D_L(z)/10~\mathrm{pc}) + k(z)
\end{equation}
that includes a $k$-correction $k(z) = -6 \log(1+z)$ in addition to
the term containing the luminosity distance $D_L(z)$. As K03 note,
this $k$-correction is negative, independent of galaxy type, and valid
for $z \lesssim 0.25$. Due to our parameterization of Hubble's
constant, we report values of $M_K - 5\log h$, but hereafter we omit
the second term for the sake of brevity.

The model for the distribution of galaxies with absolute
\mbox{$K$-band} magnitudes $M_K$, projected radii $R$, and measured
redshifts $z$ relative to a cluster at redshift $z_c$ with richness
$N_*$, velocity dispersion $\sigma_c$, and scale radius $r_c$ (or
alternatively, the probability of the cluster having a member galaxy
with these characteristics) is
\begin{align}
\label{eqn:clusterprob}
P_c &= \frac{dN}{d^2x dM_K dz} \nonumber\\
&= N_* \frac{\phi_c(M_K)}{\Phi_c(M_{K*})}
\frac{\Sigma(R)}{\sqrt{2\pi}\sigma_c(1+z_c)}
\exp\left[-\frac{c^2(z-z_c)^2}{2\sigma_c^2(1+z_c)^2}\right] ~.
\end{align}
The normalization $N_*$ is the number of cluster galaxies brighter
than $M_{K*}$ within a spherical radius $r_\mathrm{out}$ of the
cluster center. This radius ought to be roughly similar to the virial
radius. Like most previous matched filter studies, we choose
$r_\mathrm{out} = 1.0\,h^{-1}$ Mpc. Using a fixed physical radius is
convenient for calculations, and avoids adding extra scatter to
richness estimates via the use of a noisy virial radius estimate. We
discuss other possible richness measures in
Section~\ref{sec:derived}. The integrated luminosity function
$\Phi_c(M_K)$ is the spatial density of galaxies brighter than $M_K$,
and is the cumulative integral of the cluster luminosity function
$\phi_c(M_K)$. The function $\Sigma(R)$ represents the two-dimensional
spatial distribution of galaxies, and is given by a projected version
of the Navarro-Frenk-White \citep[NFW,][]{Navarro:1997p493}
profile. Finally, the Gaussian factor in redshift $z$ represents the
line of sight velocity distribution of the cluster galaxies. We
describe these components in greater detail in the following
paragraphs.

The luminosity function of cluster galaxies is assumed to be given by
a Schechter function with fixed parameters $\alpha$ and $M_{K*}$,
\begin{equation}
\label{eqn:clusterlf}
\phi_c(M_K) = 0.4\ln(10)n_*
\left(\frac{L_K}{L_{K*}}\right)^{1+\alpha}
\exp\left(-\frac{L_K}{L_{K*}}\right) ~,
\end{equation}
where $M_K-M_{K*} = -2.5\log(L_K/L_{K*})$. The integrated luminosity
function, which describes the density of cluster galaxies brighter
than a specified magnitude, is thus the incomplete Gamma function
\begin{equation}
\Phi_c(M_K) = \int_{-\infty}^{M_K} \phi_c(M) dM
= n_*\Gamma\left(1+\alpha,\frac{L_K}{L_{K*}}\right) ~.
\end{equation}
Since $\phi_c$ only appears in our calculations as a fraction of
$\Phi_c$, its normalization constant $n_*$ is unimportant. For our
first cluster finding iteration, we follow K03 in adopting the
\citet{Kochanek:2001p566} luminosity function and set $\alpha$ and
$M_{K*}$ to $-1.09$ and $-23.39$ mags, respectively. Part of the
purpose of this work is to verify the appropriateness of this choice
for the cluster galaxy luminosity function, and we update these
parameter values in our second iteration.

We model the angular distribution of cluster galaxies
as the two-dimensional projection of an NFW profile. For a
cluster with a scale radius $r_c$, this distribution, normalized by
the number of galaxies within an outer radius $C r_c$, is given in
three dimensions by
\begin{equation}
\rho(r) = \frac{1}{4 \pi r_c^3 F(C)} \frac{1}{x (1+x)^2} ~,
\end{equation}
where $x = r/r_c$ and $F(x) = \ln(1+x)-x/(1+x)$. The parameter $C$ is
similar to the concentration of the cluster. For the purpose of
normalizing this profile (or equivalently, defining $N_*$), we fix it
to $C=(1.0\,h^{-1}\,\mathrm{Mpc})/r_c$. The projected profile is
\begin{equation}
\Sigma(R) = \frac{f(R/r_c)}{2 \pi r_c^2 F(C)} ~,
\end{equation}
where
\begin{equation}
f(x) = \frac{1}{x^2-1} \left[
1 - \frac{2}{(x^2-1)^{1/2}} 
\tan^{-1} \left( \frac{x-1}{x+1} \right)^{1/2}
\right]
\end{equation}
and for $x<1$ we use the identity $-i \tan^{-1}(ix) =
\tanh^{-1}(x)$. Finally, the number of galaxies enclosed within a
circular radius $R$ is given by $N(<R) = g(R/r_c)/F(C)$, where
\begin{equation}
\label{eqn:gnfw}
g(x) = \ln\left(\frac{x}{2}\right) + 
\frac{2}{(x^2-1)^{1/2}}\tan^{-1}\left(\frac{x-1}{x+1}\right)^{1/2}
\end{equation}
\citep{Bartelmann:1996p697}. Apart from its normalization, this
density profile model has a single parameter: the core radius
$r_c$. We allow it to vary when calculating likelihoods in order to
maximize the likelihood (subject to some priors; see
Section~\ref{sec:priors}). The ``concentration'' parameter $C$ varies
with it, being defined as $(1.0\,h^{-1}\,\mathrm{Mpc})/r_c$. This is in
contrast to K03, who fix $r_c$ and $C$ to values of $0.2\,h^{-1}$ Mpc
and $4$, respectively.

The final factor in the cluster model is a Gaussian function in the
galaxy redshifts. Its parameters are the cluster redshift $z_c$ and
velocity dispersion $\sigma_c$, both of which we vary in order to
maximize the likelihood. For galaxies without a redshift measurement,
the cluster probability $P_c$ is integrated over all possible galaxy
redshifts, turning this factor into unity.

Part of the goal of this work is to tune the parameters of the matched
filter that we use to find galaxy clusters. We address the following
points:
\begin{itemize}
\item The suitability of the projected NFW profile at a range of
  scales.
\item Whether the velocity distribution of cluster galaxies is
  consistent with a Gaussian, and the richness-dependent width of the
  distribution. 
\item The luminosity function parameters $\alpha$ and $M_{K*}$, as
  well as the necessity of an extra component on the bright end to
  account for the presence of a brightest cluster galaxy (BCG)
  population.
\end{itemize}

\subsection{Derived Quantities and Richness Transformations}
\label{sec:derived}

\begin{figure}
\includegraphics[width=\columnwidth]{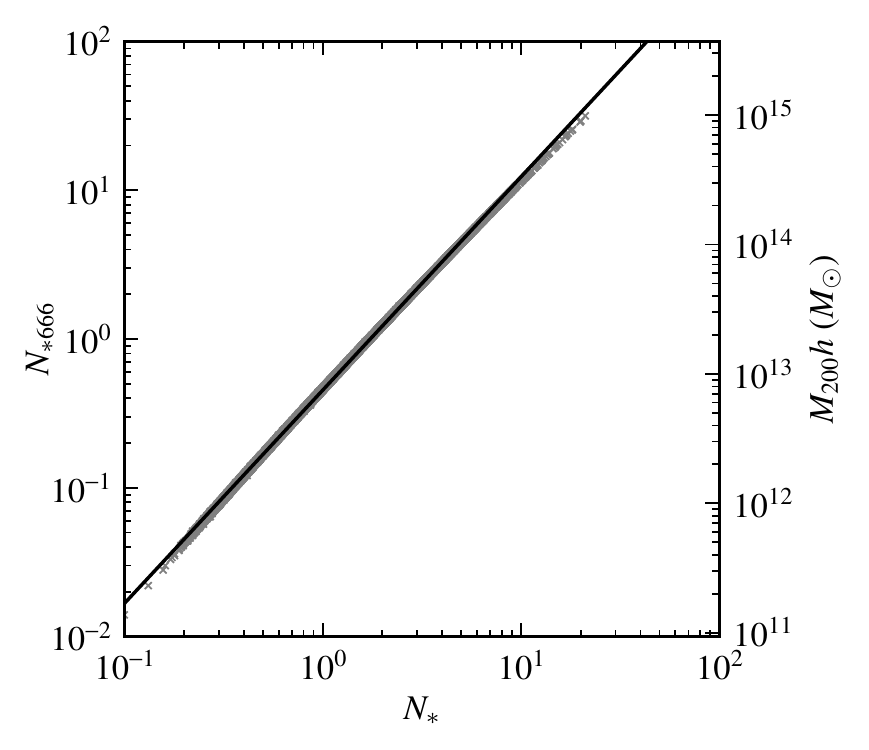}
\caption{Richness estimate $N_{*666}$ versus $N_*$ for clusters with
  $N_{*666} > 3$ and $\Delta \ln \mathcal{L} \ge 5$ found in the
  second cluster search iteration. The solid curve is the best-fit
  power law. On the right axis we show a rough estimate of the cluster
  mass $M_{200}$, as estimated from $N_{*666}$ in the X-ray stacking
  analysis of \protect{\citet{Dai:2007p917}} See
  \protect{Section~\ref{sec:derived}}.}
\label{fig:richnesses}
\end{figure}

Our primary richness measurement is $N_*$, defined as the number of
galaxies brighter than $L_*$ within a fixed radius $r < r_\mathrm{out}
= 1.0\,h^{-1}$ Mpc. Though convenient for our purposes, it is not easy
to compare to more theoretically motivated indicators, which are
usually defined with respect to a virial radius within which the
average density is some multiple of the background. To address this
issue, we define a second indicator $N_{*666}$ as the number of $L >
L_*$ galaxies within a virial radius $r_{666}$. We calculate this radius
by numerically solving the equation
\begin{equation}
N_{*\Delta_N} \equiv N_* F(r_{\Delta_N}/r_c)/F(C) = 
\frac{4\pi}{3} n_* \Delta_N r_{\Delta_N}^3 \Gamma(1+\alpha, 1)
\end{equation}
for $r_{\Delta_N}$, with the number overdensity $\Delta_N$ set to 666,
corresponding (for unit bias and $\Omega_M=0.3$) to a mass overdensity
relative to the critical density $\Delta_M = \Delta_N \Omega_M =
200$. The density $n_*$ is taken from the field galaxy luminosity
function $\phi_f$, and $\alpha$ is the slope of the cluster luminosity
function $\phi_c$. It is worth noting that aside from a relatively
unimportant dependence on $\alpha$, $N_{*666}$ is completely
determined by $N_*$ and $r_c$. The two richness indicators are highly
correlated, and their values from the second cluster search iteration
are plotted against each other in Figure~\ref{fig:richnesses}. The
best-fit relationship between them is $\log N_{*666} =
(-0.34\pm0.01) + (1.43\pm0.03)\log N_*$, and we measure a scatter
in $\log(N_{*666})$ of $0.009$ at fixed $N_*$. \citet{Dai:2007p917}
explore the relationship between $N_{*666}$ and $M_{200}$, the mass of
clusters measured within the standard radius $r_{200}$ where the
overdensity is 200 times the critical density. They find that they two
quantities are nearly linearly related, with
\begin{align}
&\log N_{*666} = \nonumber \\
&~~~~~(1.10\pm0.04) + (0.87\pm0.05)
\log\left(\frac{M_{200}h_{70}}{10^{14.6} M_\odot}\right) ~.
\end{align}
To give a rough idea of the range of masses of our cluster sample, we
show this relationship in Figure~\ref{fig:richnesses}. The plotted
values take into account their use of $h_{70}$, as well as the
difference between our second-iteration $N_{*666}$ and that of
\citet{Dai:2007p917} due to the change in the cluster luminosity
function. We discuss this difference in a few paragraphs.

The number of galaxies actually detected in a cluster of a given
richness controls its observability. This is a distance-dependent
quantity: for a nearby cluster we detect many faint galaxies that
appear brighter than our magnitude limit, while an equally rich
distant cluster will manifest only its brightest few galaxies. We
define $N_{g666}$ as the number of cluster galaxies within $r <
r_{666}$ brighter than the survey magnitude limit:
\begin{equation}
\label{eqn:expgal}
N_{g666} = N_{*666} \frac{\Gamma(1+\alpha, L_\mathrm{lim}(z_c)/L_*)}
{\Gamma(1+\alpha, 1)} ~,
\end{equation}
where $L_\mathrm{lim}(z_c)$ is the galaxy luminosity corresponding to
the survey limit at redshift $z_c$. A related quantity appears in the
first term in Equation~(\ref{eqn:likelihood}). The number of galaxies
within the \emph{circular} sampling radius $R_\mathrm{samp}$ brighter
than the limiting magnitude is given by
\begin{equation}
\label{eqn:Aexpect}
A(z) = \frac{g(R_\mathrm{samp}/r_c)}{F(C)}
\frac{\Gamma(1+\alpha, L_\mathrm{lim}(z)/L_*)}
{\Gamma(1+\alpha, 1)}
\end{equation}
for a cluster of richness $N_* = 1$, where $L_\mathrm{lim}$ is defined
as in Equation~(\ref{eqn:expgal}), and $g(x)$ is defined in
Equation~(\ref{eqn:gnfw}).

Finally, in some cases we are interested in the probability that a
certain galaxy is a cluster member. Despite its similar name, this is
not the same as the probability of finding a galaxy near a cluster
$P_c$. For a galaxy labeled $i$ near a cluster labeled $k$, we define
its membership probability as
\begin{equation}
\label{eqn:membprob}
P_\mathrm{memb}(i,k) = \frac{P_c(i,k)}{\sum_j P_c(i,j) + P_f(i)} ~,
\end{equation}
where $P_f$ and $P_c$ are defined in Equations~(\ref{eqn:fieldprob})
and (\ref{eqn:clusterprob}), the and sum is over all clusters. This
definition is nearly identical to that of \citet{Rozo:2009p601},
except that they consider only the $j=k$ term of the sum.

There is a subtle but important difference between the values of $N_*$
and $N_{*666}$ that we find in our first cluster iteration and those
we find in the second iteration. This is because we change the
parameters of the luminosity function between the iterations, most
importantly the cutoff magnitude $M_{K*}$. Since $N_*$ is defined as
the number of galaxies brighter than $M_{K*}$, we change this count
when we change $M_{K*}$. In Section~\ref{sec:lumfunc} we discuss the
change in the matched filter luminosity function between the first and
second search iterations. As the cluster search algorithm varies $N_*$
to maximize the likelihood, it is effectively fitting the matched
filter luminosity function to the observed galaxy distribution by
varying its normalization. Therefore, our best estimate of the
transformation between $N_{*1}$ from the first cluster search
iteration and $N_{*2}$ from the second is the ratio of normalizations
of the Schechter functions that best fit the stacked cluster
luminosity function. We perform these fits in
Section~\ref{sec:lumfunc}. The normalization of the second-iteration
luminosity function is a factor of 0.838 lower than that of the first
iteration; therefore, $N_{*2} \approx 0.838 N_{*1}$ for any given
cluster. With the (reasonable) assumption that $r_c$ does not depend
on the shape of the luminosity function, the virial richness
$N_{*666}$ varies the same way. The transformation of $N_{g666}$ is
more complicated; it involves not only the ratio of normalizations but
the fraction of galaxies brighter than the survey limit. The ratio of
the second-iteration $N_{g666}$ to the first-iteration one is 
\begin{displaymath}
0.838 \frac{\Gamma(1+\alpha_2, L_\mathrm{lim}/L_{*2})}
{\Gamma(1+\alpha_1, L_\mathrm{lim}/L_{*1})}
\frac{\Gamma(1+\alpha_1,1)}{\Gamma(1+\alpha_2,1)} ~,
\end{displaymath}
where the subscripts 1 and 2 refer to the first and second iterations,
and $L_\mathrm{lim}$ is the limiting luminosity as in
Equation~(\ref{eqn:expgal}). This value is redshift-dependent, varying
from 0.868 at $z_c = 0.01$ to 1.328 at $z_c = 0.1$. It is unity at a
redshift of 0.047, and since this is a fairly typical redshift for our
clusters we adopt this value, so that there is no change in $N_{g666}$
between iterations. We tested these ratios by matching a subset of our
second-iteration cluster catalog to our first-iteration catalog, and
found them to be correct predictions. Throughout this paper, we report
the values of $N_*$ and $N_{*666}$ calculated by our
algorithm without applying the correction factor, but we caution
the reader to take care in comparing their values. In some cases we
use the factor to define subsets of the cluster catalogs that should
be roughly equivalent between iterations, and we note our use of the
transformation factors in those instances.

\section{Characteristics of Stacked Clusters}
\label{sec:stacking}

We cannot characterize the radial profile, luminosity function, or
velocity distribution of individual clusters in any detail, because
the typical cluster contains too few galaxies. But by ``stacking''
large number of clusters, we can determine their average
properties. Methods similar to this have been used in several studies
\citep[e.g.,][]{Carlberg:1996p32, Dai:2007p917, Rykoff:2008p1106,
Rozo:2010p645}. We apply the process to a sample of clusters with
likelihoods $\Delta \ln \mathcal{L} > 5$ and expected galaxy number
$N_{g666} > 3$. K03 found that this last requirement resulted in an
extremely pure sample of clusters.

The stacking process consists of selecting the galaxies within some
physical radius of the cluster center and constructing a histogram of
the relevent quantity (i.e., radial position for a radial profile or
absolute magnitude for a luminosity function), subtracting an
appropriate background (estimated using the entire galaxy sample), and
averaging the results for the sample of clusters. We estimate the
uncertainties in our averages using bootstrap resampling, an
approach which we advocate for future studies. We resample both the
cluster {\em and} galaxy lists with replacement to construct the
bootstrap uncertainties. In all cases we resample 100 times.

Before stacking, we separate our cluster catalog into richness
bins. We divide the space between $N_{*666} = 0.1$ and $N_{*666} = 30$
into five logarithmic bins 0.5 dex wide. For the first cluster catalog
(i.e., the catalog resulting from the first cluster search iteration),
this division excludes five clusters with $N_{*666} < 0.1$ and two
with $N_{*666} > 30$, leaving a total of membership of 58, 365, 596,
425, and 85 clusters in the five bins, in increasing order of
richness. In the second cluster catalog, we multiply the bin edges by
a factor of 0.838 to account for the fact that the richness
measurements are systematically lower than they were in the first
iteration (see Section~\ref{sec:derived}). These five bins contain 46,
450, 867, 594, and 125 clusters. We further subdivide these bins into
membership samples with $3 \le N_{g666} < 5$, $5 \le N_{g666} < 7$, $7
\le N_{g666} < 10$, and $N_{g666} \ge 10$ detected galaxies. This is
to check whether there is any bias in our estimates of the cluster
properties with the number of detected galaxies. Since the number of
detected galaxies depends primarily upon redshift, changes between
these subsamples could also indicate evolution of average cluster
properties, but at the low redshifts of our clusters we expect little
evolution.

It is desirable to use a richness estimator with a small scatter
relative to the actual cluster richness \citep[e.g.,][]{Rozo:2009p601,
  Rykoff:2011p2089}. Although $N_{*666}$ is useful for comparison with
theoretical studies, it is possible that the nature of its definition
(i.e., with respect to other noisy quantities) makes it a noisier
estimator than the more simply defined $N_*$. To check this, we divide
the clusters resulting from our second cluster search iteration into
bins of $N_*$, matching the bin edges to those of our previous bins
using the best-fit relation between $N_*$ and $N_{*666}$ described in
Section~\ref{sec:derived}. We examine the uncertainties in our
average profiles, velocity distributions, and luminosity functions
resulting from the bootstrap resampling, comparing them to those from
the $N_{*666}$ binned samples. There are no significant differences
in the size of the error bars, so we cannot conclude on those grounds
that $N_*$ is a better richness estimator than $N_{*666}$. Given the
extremely tight correlation between $N_*$ and $N_{*666}$, this is not
surprising; the richness bins contain nearly identical sets of
clusters in both cases.

\subsection{Density Profile}
\label{sec:profile}

We first compare the average surface density distribution $\Sigma(R)$
to the projected NFW we use for the matched filter. We measure the
density profiles in a series of logarithmically spaced annuli with
inner and outer edges $R_{\mathrm{in},j} < R_j < R_{\mathrm{out},j}$,
centered on $R_j = (R_{\mathrm{in},j} R_{\mathrm{out},j})^{1/2}$, and
with area $A_j = \pi (R_{\mathrm{out},j}^2 -
R_{\mathrm{in},j}^2)$. The expected number of background galaxies in
an annulus is
\begin{align}
b_j &= 2 \pi B \nonumber
\left[ \left( 1+R_{\mathrm{in},j}^2/D_A(z)^2 \right)^{-1/2} -
\left( 1+R_{\mathrm{out},j}^2/D_A(z)^2 \right)^{-1/2} \right] \\
&\approx B A_j/D_A(z)^2 ~,
\end{align}
where $B$ is the angular surface density of galaxies to the survey
magnitude limit (by ``background'', we mean a combination of
foreground and background galaxies). The higher-order terms in the
curved-sky Taylor expansion become important when we examine nearby
clusters at large radii. If we count $N_j$ galaxies within the $j$-th
annulus, the surface density profile of the cluster, scaled by its
richness $N_*$, is
\begin{equation}
\Sigma(R_j) = \frac{1}{N_*} \frac {N_{*666}}{N_{g666}}
\left[ \frac{N_j-b_j}{A_j} \right] ~.
\end{equation}
The quantity $(N_j-b_j)/A_j$ is the surface number density of cluster
galaxies, after subtracting the mean background $b_j$. The factor
$N_{*666}/N_{g666}$ converts the observed number of galaxies (brighter
than the survey magnitude limit) to the average number of $L>L_*$
galaxies; see Equation~(\ref{eqn:expgal}). We average this estimate of
the surface density over all clusters in each richness and membership
bin. At small and medium scales, all the clusters in the sample
contribute, but at the largest scales the number of contributing
clusters declines slightly because a few clusters overlap the Galactic
latitude boundary $|b| > 6.0^\circ$. We also calculate the average
$N_*$ and $r_c$ for the clusters in each richness bin.

\begin{figure}
\includegraphics[width=\columnwidth]{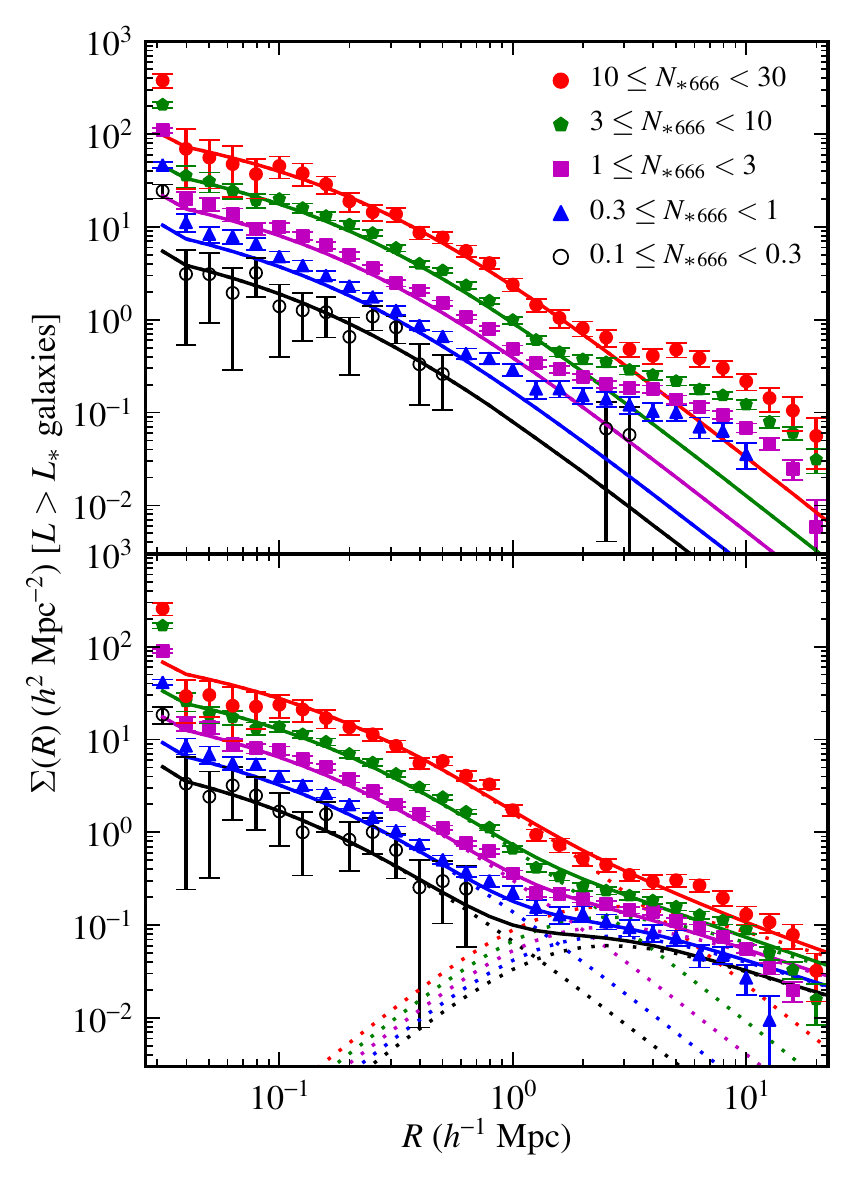}
\caption{The projected distributions of galaxies for clusters from the
  first (top panel) and second (bottom panel) cluster search
  iterations. They are divided into richness classes from low (open
  circles) to high (filled circles) richness; note that the second
  iteration bin edges are smaller than those of the first iteration
  (see \protect{Section~\ref{sec:derived}}). This figure uses the full
  sample of clusters with $N_{g666} > 3$ (i.e., the union of the
  membership classes). The points are the observed distributions, with
  their bootstrap uncertainties. The central density spike is an
  artifact resulting from our practice of centering every cluster on a
  galaxy. The first panel shows the projected NFW model. In the second
  panel, the NFW and two-halo profiles are shown in dotted lines,
  while their sum is the solid line. Only the NFW component was used
  in the matched filter, which extends to $1.0\,h^{-1}$ Mpc. Note that
  we can measure the cluster profile out to 20 Mpc from the center of
  all but the poorest clusters. The mean background of unrelated
  galaxies has been subtracted.}
\label{fig:profile}
\end{figure}

Figure~\ref{fig:profile} shows the projected density profiles of the
stacked clusters from the first and second cluster search iterations
for each richness bin, including all clusters with at least $N_{g666}
\ge 3$ member galaxies and $\Delta \ln \mathcal{L} \ge 5$. In each bin, the
profile has been multiplied by the average $N_*$ of the bin. We
superpose the projected NFW profile $\Sigma(R)$ that was used to find
the clusters, calculated using the average $N_*$ and $r_c$ in each
bin. In the second panel we also show an ad-hoc profile for the
two-halo contribution suggested by the observed average profiles (more
on that in a few paragraphs). It is important to note that the model
profiles are not fits to the galaxy profiles, but are simply the
matched filter model calculated using the average cluster
properties. We note three distinct radial regimes in the profiles.

First, at small radii ($R \lesssim 0.1\,h^{-1}$ Mpc) the profiles are
sensitive to the method used to center the clusters. The difficulty of
determining the central galaxy density profile of clusters due to the
method used to center the cluster is well known
\citep[e.g.,][]{Beers:1986p557}. The most obvious sign we see is a
distinct density spike in the innermost bin; this is due to our
practice of always centering a cluster on a galaxy. We avoid
``smearing'' the contribution of the central galaxy over the region
with $R<r_c$, but there will always be problems reconstructing a
possibly singular average central density profile in the presence of
the shot noise from the galaxies sampling the profile. We experimented
with using a cluster position estimated by averaging the membership
probability weighted positions of the cluster galaxies. While this
eliminated the central spike, the profile shape began to depend on the
number of member galaxies $N_{g666}$. Essentially, we measure the true
average profile convolved with the position measurement errors, and
these increase considerably as the number of member galaxies available
for the average decreases. This problem has been seen by
\citet{Dai:2007p917} and \citet{Rykoff:2008p1106}, who both stack
X-ray images of optically-selection clusters. Despite these problems,
our profiles match the expected shape reasonably well apart from the
innermost radial bin.

Second, on intermediate scales ($0.1\,h^{-1}$ Mpc $\lesssim R \lesssim
1.6\,h^{-1}$ Mpc), which dominate our detection and parameter
estimation, the average surface density of the galaxies matches
reasonbly well the profile shape expected from our matched filter. The
observed and NFW surface density profiles have formal chi-square
differences, for 13 degrees of freedom, of 9.0 to 46 for the first
iteration and 2.8 to 42 for the second iteration. These large values
arise largely from $\sim 10\%$ normalization differences between the
data and the model (which was calculated with no free parameters);
these differences are unimportant for the matched filter, where the
normalization is free to vary.

Third, on large scales ($R \gtrsim 1.6\,h^{-1}$ Mpc), the observed
surface density lies above the projected NFW profile. The
NFW model represents only the virialized regions of the cluster
and neglects the correlated structure outside the virial radius ---
the ``two-halo'' term, in the parlance of the halo model of large
scale structure \citep[e.g.,][]{cooray:2002p1}. This outer region of
clusters should provide a good testing ground for the halo model
because it represents the dividing line between linear and nonlinear
regimes. We create a crude model for the profile of the two-halo term
consisting of a broken power law:
\begin{equation}
\Sigma_\mathrm{2h}(R) = \frac{a_\mathrm{2h}(N_*)
\left( R/R_\mathrm{2h} \right)^{\beta_\mathrm{in}}}
{\left( 1+ R^2/R_\mathrm{2h}^2
  \right)^{(\beta_\mathrm{in}+\beta_\mathrm{out})/2}} ~,
\end{equation}
where $a_\mathrm{2h}(N_*)$ is a richness-dependent normalization
constant for the two-halo term and $R_\mathrm{2h}$ is a break
radius. The inner and outer slopes $\beta_\mathrm{in}$ and
$\beta_\mathrm{out}$ we set to 2 and 0.8 respectively; the former
because it made the enclosed mass an analytic function, and the latter
by analogy with the slope of typical correlation functions on these
scales. We adjust the normalization and scale radius until they
roughly match the observed second-iteration profile, finding that
$a_\mathrm{2h}(N_*) = (0.17\pm0.01)
N_*^{(0.33\pm0.04)}\,h^2\,\mathrm{Mpc}^{-2}$ and that $R_\mathrm{2h} =
(1.66\pm0.10)\,h^{-1}$ Mpc. We do not add the two-halo term to the
matched filter in the second cluster search iteration, because its
contribution inside $1.0\,h^{-1}$ Mpc (the sampling radius) is very
small relative to the NFW component.

\begin{figure}
\centering
\includegraphics[width=0.92\columnwidth]{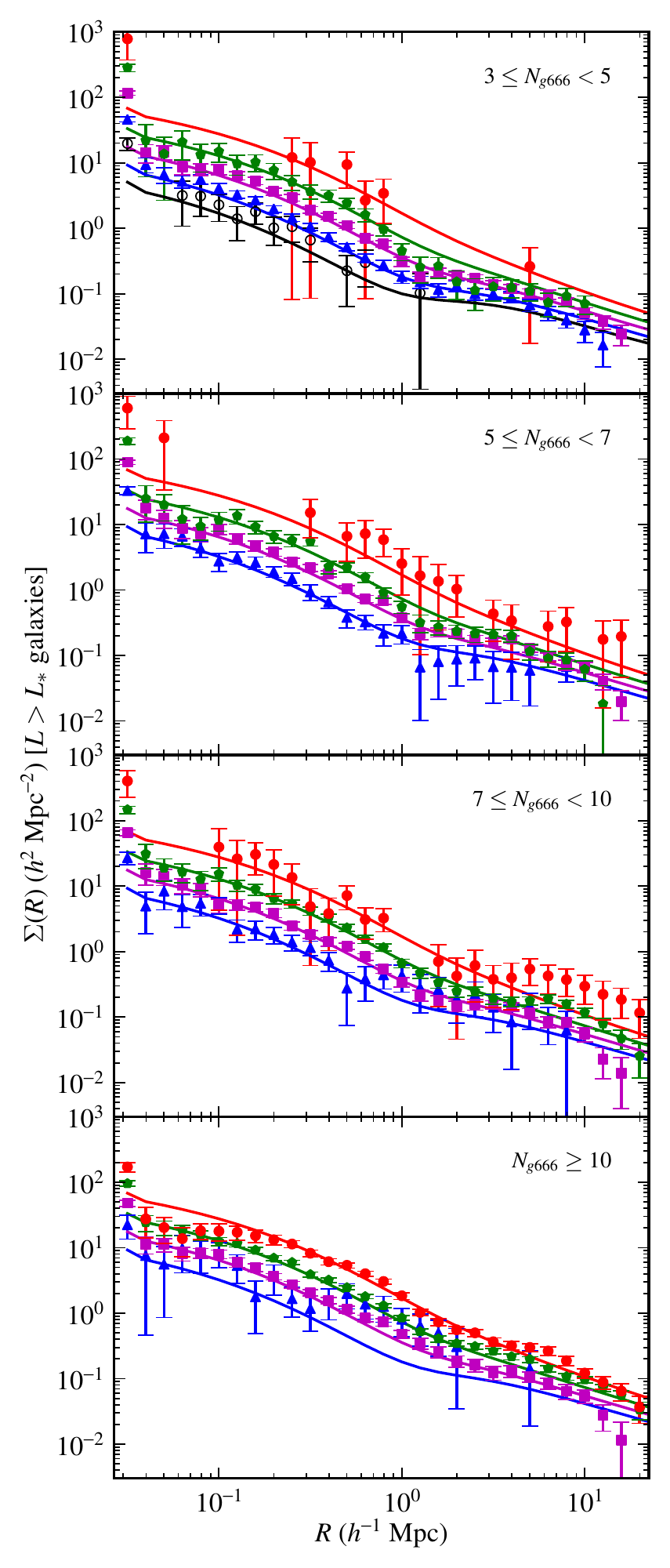}
\caption{Projected cluster profiles for cluster subsamples with
  different numbers of detectable galaxies $N_{g666}$. The clusters
  with the fewest visible galaxies are in the top panel, and the
  number increases toward the bottom panel. The points' colors and
  shapes indicate richness bin, as in
  \protect{Figure~\ref{fig:profile}}, and the solid curves show the
  sum of the NFW and two-halo term. We do not see strong evidence for
  evolution in the profile with $N_{g666}$.}
\label{fig:profile_g}
\end{figure}

For second-iteration clusters, we repeat the stacking process for the
membership subsamples of each richness bin in order to check for
evolution of the cluster profiles with the number of detected
galaxies. The resulting richness estimates are shown in
Figure~\ref{fig:profile_g}. We plot only samples where the product of
the number of clusters and the average value of $N_{g666}$ is greater
than 30; this excludes three low-richness samples. On scales less than
$\sim 1\,h^{-1}$~Mpc, we do not see any evidence for changes in the
profile with the number of visible galaxies, unless the profile is
very noisy.

\subsection{Velocity Distribution}
\label{sec:sigma}

Our cluster search algorithm produces two estimates of the velocity
dispersion of each cluster. The first estimate is the value of
$\sigma_c$ used in the matched filter. We maximize the likelihood for
each cluster by varying the velocity dispersion using MCMC
optimization, subject to our priors. We also compute the velocity
dispersion using galaxies with redshift measurements within $Cr_c$ of
the center of each cluster, weighting the contribution of each galaxy
by its cluster membership probability $P_\mathrm{memb}$ (see
Equation~(\ref{eqn:membprob})). We average the latter velocity
dispersion estimate in each richness bin, and use this average to
define our model in the following analysis.

As with the cluster density profiles, the typical cluster contains too
few galaxies to accurately estimate the velocity distribution in a
non-parametric way. So we again stack large numbers of clusters in
fixed richness bins to determine the average velocity distribution. We
first identify all galaxies with measured redshifts within the
projected virial radius, $R \le r_{666}$, of each cluster, so as to
compare velocity histograms inside the virialized region for clusters
of differing richness. We discard those galaxies which lack redshift
measurements; this should be relatively unbiased with respect to the
velocity distribution since any target selection method for measuring
redshifts is unbiased with respect to the relevant velocity
differences. We then construct histograms of the rest frame line of
sight velocities (relative to the systemic velocity) $\Delta v =
(v-cz_c)/(1+z_c)$, considering only velocities where $|\Delta v| \le
2000$ km s$^{-1}$. We use variable bin widths of 50,
75, 100, 150, and 250 km s$^{-1}$ for the five richness bins. Finally,
we average the histograms over the clusters in each richness bin,
excluding clusters with $N_v < 5$ galaxies with measured redshifts and
clusters with virial radii extending into the region where $|b| <
6^\circ$. The requirement of five redshift measurements limits the
effect of the sample variance bias. We also calculate the average
richness and velocity dispersion of the clusters in each richness bin,
weighted by $N_v - 1$.

\begin{figure}
\includegraphics[width=\columnwidth]{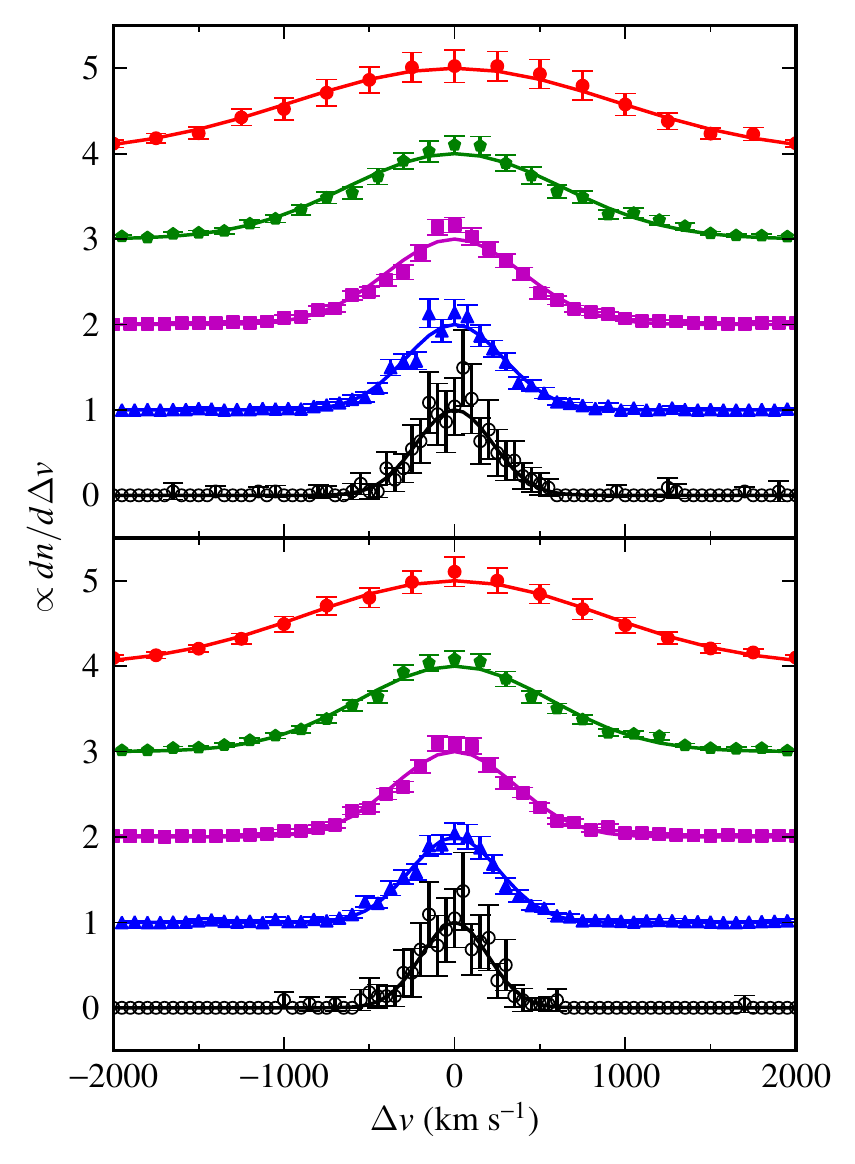}
\caption{The rest-frame line of sight velocity distribution of the
  first-iteration (first panel) and second-iteration (second panel)
  cluster galaxies in the five richness bins. The points show the
  average velocity distributions, with the same colors and shapes as
  in \protect{Figure~\ref{fig:profile}}, and the solid curves are the
  Gaussian models used by the matched filter.}
\label{fig:sigma}
\end{figure}

\begin{figure}
\centering
\includegraphics[width=0.92\columnwidth]{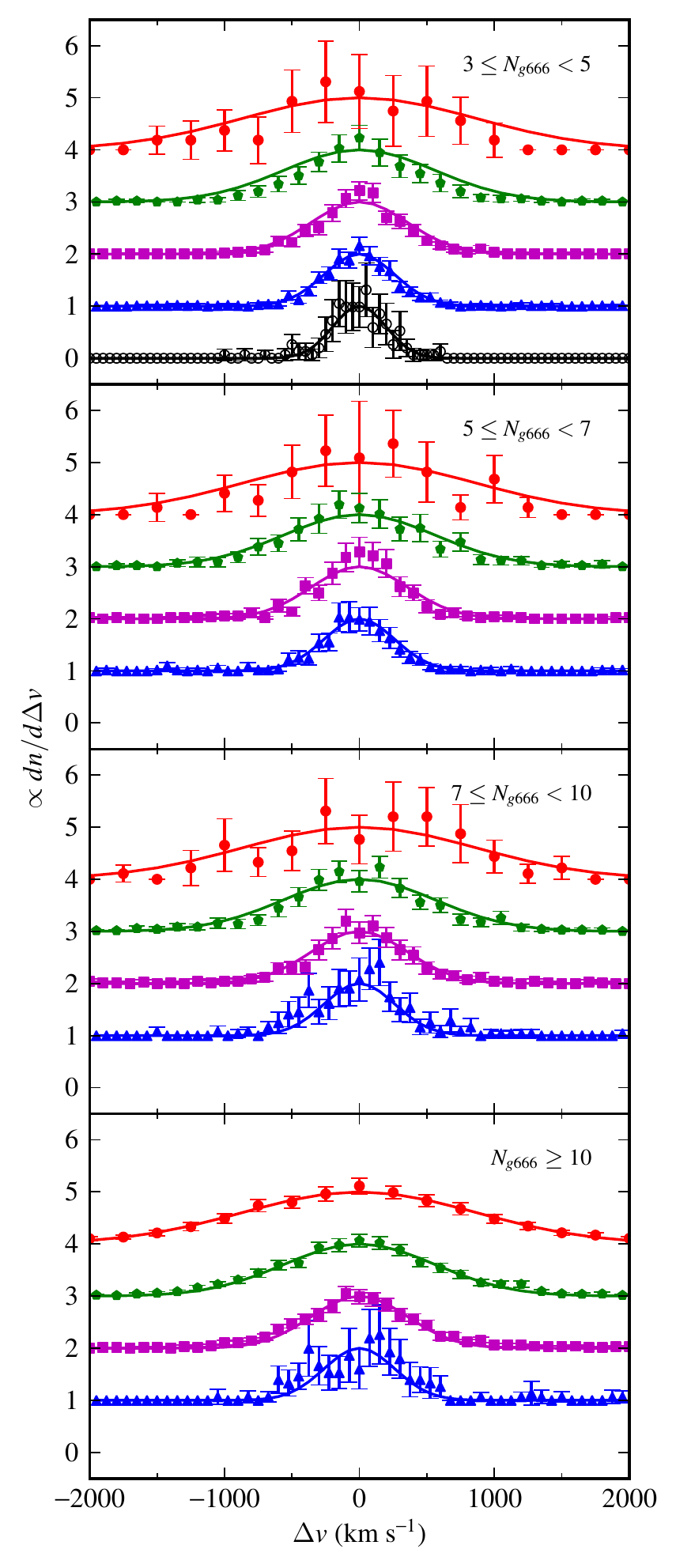}
\caption{Velocity distributions for cluster subsamples with different
  numbers of detectable galaxies $N_{g666}$, which increases from the
  first frame to the last. The points' colors and shapes indicate
  richness bin, as in \protect{Figure~\ref{fig:sigma}}, and the solid
  curves are identical to those in that figure's second panel. We see
  no strong evidence for evolution of the velocity distributions with
  $N_{g666}$.}
\label{fig:sigma_g}
\end{figure}

Figure~\ref{fig:sigma} shows the velocity distributions measured in
this way for the first and second cluster search iterations, for the
same richness bins we used for our profile measurement. Superposed on
the distributions are Gaussian curves with widths set by the average
velocity dispersion in the bins. The normalization of the curves is
arbitrary, and they are adjusted to match the observed
distributions. We find excellent agreement between the predicted curve
and the observed distribution. Figure~\ref{fig:sigma_g} shows the
velocity distributions as a function of visible galaxy number
$N_{g666}$. We impose the same cut as was previously used to weed out
three low-richness samples. No significant evolution of the velocity
distributions with the number of detected galaxies $N_{g666}$ can be
seen. Although the velocity distributions look somewhat narrow in the
lowest-membership sample, we think this is due to the low number of
galaxies per cluster.

One advantage of the stacked clusters is that they contain many
galaxies, so it is easy to estimate a Gaussian width without using a
statistical method that is dominated by the tails of the
distribution. We clip the velocity distributions at twice the average
velocity dispersion for each richness bin. We then sort the velocity
differences and estimate the velocity dispersion as one-half the
velocity range encompassing 68.3\% of the galaxies centered on the
median. Like a simple velocity dispersion, this estimate is exact for
a Gaussian distribution, but it is insensitive to interloping field
galaxies. The resulting velocity dispersion estimates are reported,
along with the dispersions determined by averaging over each richness
bin, in Table~\ref{tab:veldisp}. The differences between the two
estimates are less than 10\%.

\begin{deluxetable}{lcc}
  \tablewidth{0pt}
  \tablecaption{Velocity Dispersions
    \label{tab:veldisp}}
  \tablehead{
    \colhead{Richness\tablenotemark{a}} &
    \colhead{$\langle \sigma_c \rangle$ (km s$^{-1}$)\tablenotemark{b}} &
    \colhead{$\sigma_\mathrm{sort}$ (km s$^{-1}$)\tablenotemark{c}}
  }
  \startdata
  $0.1 \le N_{*666} < 0.3$         & $197$ & $190$ \\
  $0.3 \le N_{*666} < 1$           & $261$ & $250$ \\
  $\phn\phd1 \le N_{*666} < 3$     & $357$ & $338$ \\
  $\phn\phd3 \le N_{*666} < 10$    & $561$ & $531$ \\
  $\phd10 \le N_{*666} < 30$       & $862$ & $821$
  \enddata
  \tablenotetext{a}{Though we do not show it here, the bin edges have
  been reduced by a factor of 0.838 because $N_{*666}$ is
  systematically smaller for the second cluster search iteration (see
  \protect{Section~\ref{sec:derived}}).}
  \tablenotetext{b}{Average velocity dispersion in the bin.}
  \tablenotetext{c}{Velocity dispersion estimated by sorting the
  velocities (see \protect{Section~\ref{sec:sigma}}).}
\end{deluxetable}

\subsection{Luminosity Function}
\label{sec:lumfunc}

The final distribution we consider is the luminosity function of the
cluster galaxies. Since clusters at different distances have different
galaxy luminosities corresponding to the survey magnitude limit, some
care is required in combining clusters within a richness bin. We first
construct a histogram of the absolute magnitudes of galaxies within
$R_\mathrm{samp} = 1.0\,h^{-1}$ Mpc of each cluster center, with bins
$1/3$ mag in width. Each histogram extends to the faintest bin whose
faint edge is brighter than the survey limit. In this same set of
magnitude bins we calculate the ``background'' of interloping galaxies
by taking the distribution of {\em apparent} magnitudes of our galaxy
sample, shifting this distribution by the distance modulus
$\mathcal{D}(z_c)$, and scaling the resulting distribution by the
ratio of the angular area of the cluster to that of the whole
survey. We subtract this background from the luminosity function and
scale the difference by the factor $N_* g(R_\mathrm{samp}/r_c)/F(C)$
to account for our cylindrical sampling volume. Finally, we average
the result over all the clusters in each richness bin, weighting the
clusters by this same factor. The number of clusters contributing to
the luminosity function estimate is a strong function of absolute
magnitude; the faintest bins only have contributions from the nearest
clusters, while the brightest bins average over tens to hundreds of
clusters, depending on the richness and membership bin. As before, we
exclude clusters with sampling radii extending into the excluded
region of low Galactic latitude where $|b| < 6^\circ$.

\begin{figure}
\includegraphics[width=\columnwidth]{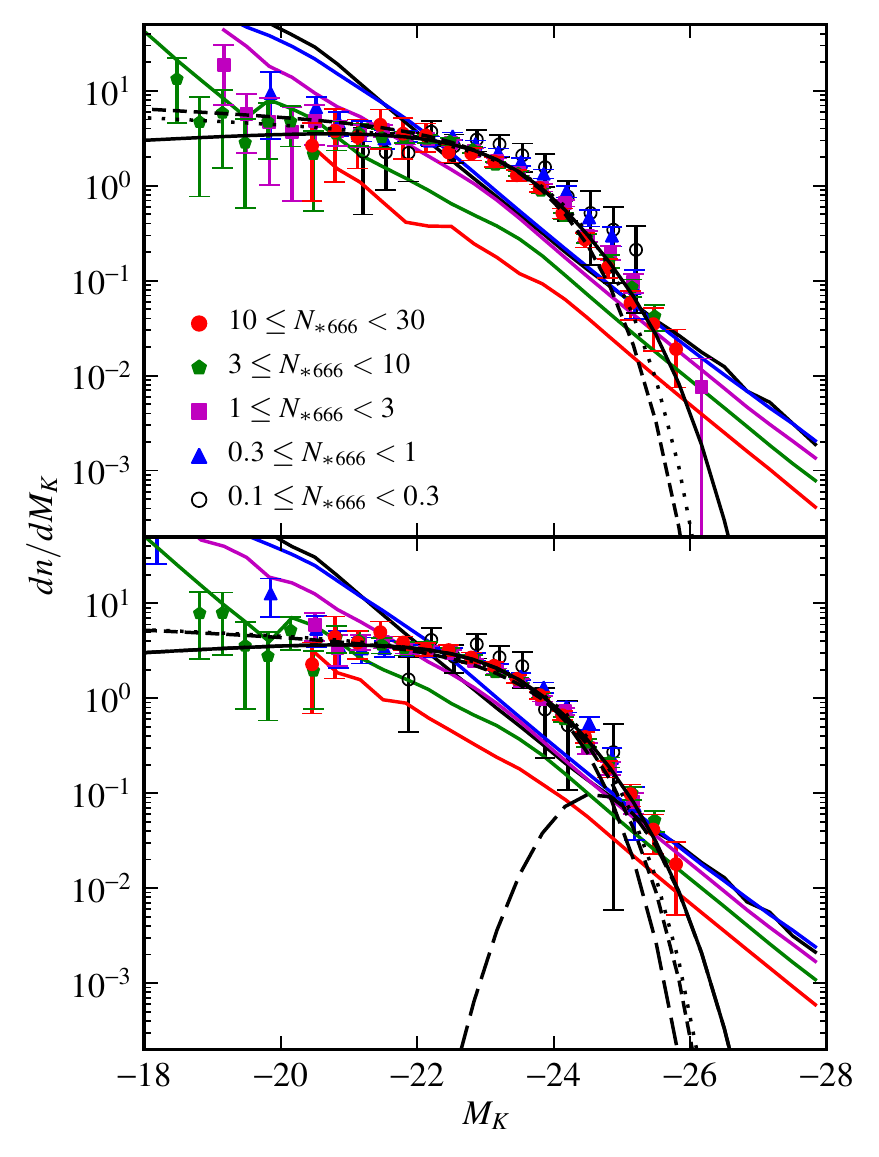}
\caption{The cluster luminosity function for first-iteration (first
  panel) and second-iteration (second panel) clusters in the five
  richness bins. The points show the measured luminosity functions,
  normalized to unit $N_*$. In each panel, the short-dashed curve
  signifies the Schechter function used in the matched filter, and the
  dotted curve is the best-fit Schechter function. The solid curves
  indicate the best-fit sum of Schechter and BCG components, which are
  also shown separately as long-dashed lines in the second panel. The slanted
  lines show the level of foreground/background contamination that
  is subtracted to make the estimate of the luminosity function
  for each richness class; the contrast of clusters relative to this
  contamination depends strongly on richness. At very high and very
  low luminosities, the foreground/background subtraction fails.}
\label{fig:lumfunc}
\end{figure}

Figure~\ref{fig:lumfunc} shows the observed cluster galaxy luminosity
function for clusters from our first and second search iterations. In
each case, the luminosity function used for the matched filter is
plotted as a short-dashed curve, with a normalization adjusted to best
fit the observed data. The curve in the first panel clearly lies below
the data, especially at the bright end. So we fit a second Schechter
function to the data; this is plotted as a dotted curve. This function
was characterized by $M_{K*} = -23.59\pm0.05$ mags and $\alpha =
-1.08\pm0.03$. We used this function for the second matched filter; it
is therefore the short-dashed curve in the second panel. It closely
matches the best-fit Schechter function for that panel, which has
parameters $M_{K*} = -23.64\pm0.05$ mags and $\alpha =
-1.07\pm0.03$. This fit is good, producing a chi-square of 169.6 for
130 degrees of freedom.

There has been considerable interest in the luminosity fuction of BCGs
and whether they lie on the same Schechter function as satellite
galaxies \citep[e.g.,][]{Yang:2008p248}. In Figure~\ref{fig:lumfunc}
we see a small excess at the bright end of the luminosity function
relative to the best-fit (dotted) Schechter function, suggesting that
we are seeing a separate population of BCGs. We parameterize this
population using a Gaussian luminosity distribution in magnitude
space. To investigate the BCG population, we attempt to identify in
each cluster the galaxy with the greatest probability of being the
BCG. The BCG probability is the product of the galaxy's cluster
membership probability, defined in Equation~(\ref{eqn:membprob}), and
the probability that the galaxy's luminosity is drawn from the
Gaussian BCG distribution. For the clusters found in the first
iteration, we made a guess at this distribution, centering the
Gaussian at $-25.1$ mags, with a variance of
$(0.4~\mathrm{mags})^2$. The exact parameters of this distribution are
not very important; its main function is to help us rank the galaxies
by BCG probability. We exclude any galaxy that lacks a redshift
measurement, has membership probability $P_\mathrm{memb}<0.5$, or is
farther than $R_\mathrm{samp} = 1.0\,h^{-1}$ Mpc from the cluster
center. Out of our sample of 1532 clusters with $N_{g666} > 3$, this
process yields BCGs for 1507. The magnitude distribution of these
galaxies is close to Gaussian in shape, and has mean $-24.61$ mags and
variance $(0.56~\mathrm{mags})^2$. We adopt these parameters for the
BCG luminosity distribution, and repeat our fit of the luminosity
function with this Gaussian term included. We find best fit values of
$M_{K*}=-23.22\pm0.07$ and $\alpha=-0.90\pm0.05$, with the BCG
distribution peaking at $(0.045\pm0.003) \phi_c(M_{K*})$. This
two-part luminosity function is shown as a solid curve in the first
panel of Figure~\ref{fig:lumfunc}. The best-fit luminosity function
for the second-iteration cluster sample is nearly identical, with
$M_{K*} = -23.27\pm0.07$ mags and $\alpha = -0.89\pm0.05$. The BCG
curve is centered on $-24.57$ mags, with a variance of
$(0.59~\mathrm{mags})^2$, and reaches a maximum of $(0.048\pm0.009)
\phi_c(M_{K*})$. This function is likewise plotted as a solid curve in
the second panel of Figure~\ref{fig:lumfunc}, and its components are
shown as long-dashed curves. This fit is slightly better than the
Schechter-only fit, with a chi-square of 141.6 for 129 degrees of
freedom.

\begin{figure}
\centering
\includegraphics[width=0.92\columnwidth]{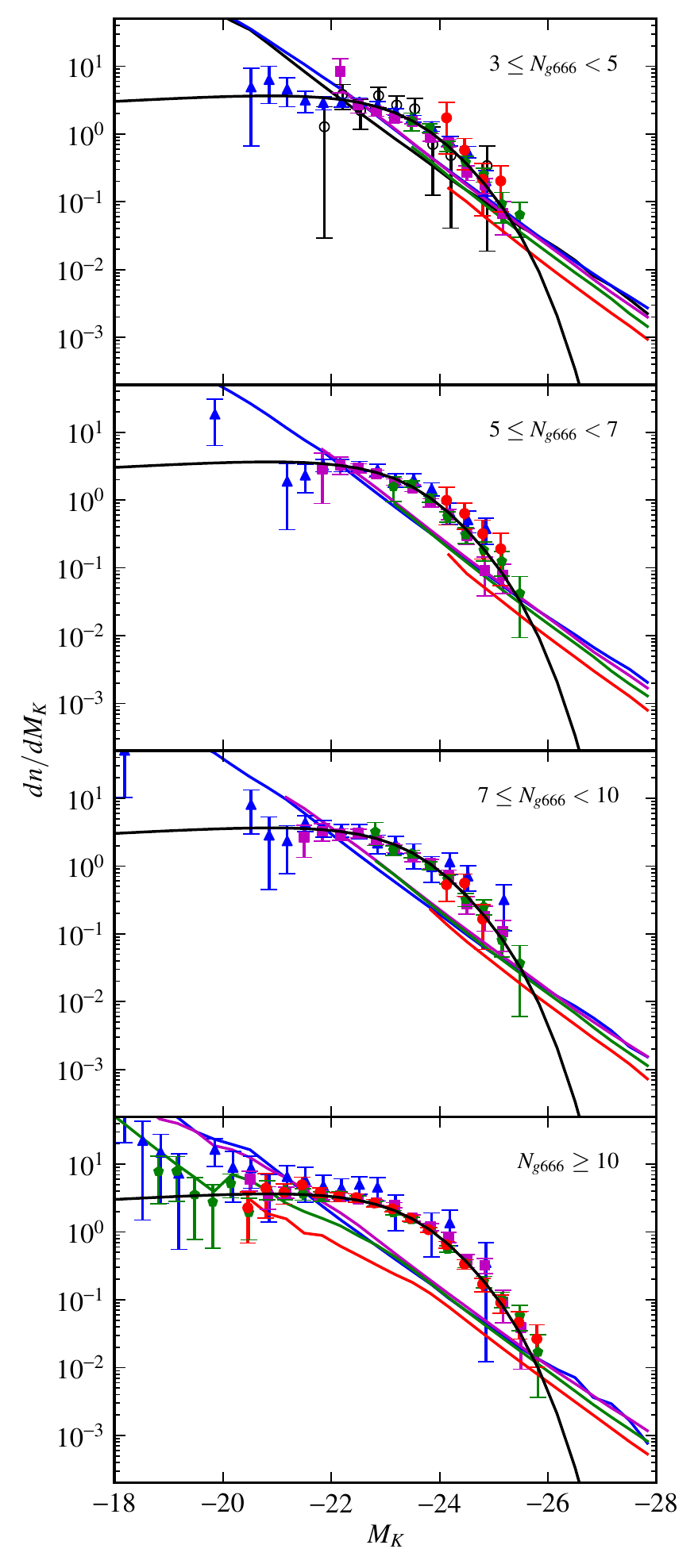}
\caption{Cluster luminosity function for cluster subsamples with
  different numbers of detectable galaxies $N_{g666}$, which increases
  from the first frame to the last. The points' colors and shapes
  indicate richness bin, as in \protect{Figure~\ref{fig:lumfunc}}, and
  the solid curves are identical to those in that figure's second
  panel. We see no strong evidence for evolution of the luminosity
  function with $N_{g666}$.}
\label{fig:lumfunc_g}
\end{figure}

The luminosity function calculated using the membership subsamples is
shown in Figure~\ref{fig:lumfunc_g}. In each panel we also plot a
solid curve showing the same best-fit luminosity function as is shown
in the second panel of Figure~\ref{fig:lumfunc}. We again see no
strong evidence of evolution with the number of observed cluster
members.

Our luminosity function results broadly agree with previous estimates
for the IR luminosity functions of cluster galaxies.  For example,
\citet{Balogh:2001p117} estimate the \mbox{$K$-band} luminosity
function using 2MASS galaxies matched to groups and clusters in the
LCRS. For groups ($\sigma < 400$ km/s, or $N_{*666} \lesssim 3$), they
find estimates of $M_{K*} = -23.58 \pm 0.13$ and $\alpha = -1.14 \pm
0.26$, whereas for clusters ($\sigma > 400$ km/s, or $N_{*666} \gtrsim
3$) they find $M_{K*} = -23.81 \pm 0.40$ and $\alpha = -1.30 \pm
0.43$. Similarly, \citet{Lin:2004p745} examine X-ray selected clusters
in 2MASS data, finding $-1.1 \lesssim \alpha \lesssim -0.84$ and
$-24.57 \le M_{K*} \le -23.25$. In their study of clusters from the
Canadian Network for Observational Cosmology survey
\citep{Yee:1996p269} at a median redshift of $z=0.3$,
\citet{Muzzin:2007p1106} find $M_{K*} = -23.76\pm0.15$ and $\alpha =
-0.84\pm0.08$. We have adjusted the values of $M_{K*}$ from
\citet{Lin:2004p745} and \citet{Muzzin:2007p1106} to account for their
use of $h = 0.7$.

\section{Priors}
\label{sec:priors}

As alluded to in previous sections, we make a number of adjustments to
the likelihood function in Equation~(\ref{eqn:likelihood}) in order to
stabilize parameter estimation and reflect prior knowledge. Each
modification takes the form of an additive term, usually the logarithm
of a multiplicative prior probability distribution. The exact formulas
for these additions, labeled Prior~I through Prior~VI, are listed in
Table~\ref{tab:priors}, with their parameters listed in
Table~\ref{tab:priorparams}. The first prior imposes a prior on the
richness $N_*$ with a slope of $-2$, and turning over at $N_*=0.1$,
the expected richness of the poorest groups. This enforces a
reasonable mass function for clusters, with a power-law
slope. Prior~II makes the likelihood independent of $\sigma_c$ when
the candidate cluster contains only one galaxy with a redshift
measurement, removing a bias in the optimization of this parameter's
value. Prior~III reflects empirical relationships between $N_*$ and
the parameters $\sigma_c$ and $r_c$, imposing a Gaussian centered on a
power-law relationship between the richness and the parameter, with a
width designed to match the scatter in the parameter at fixed
richness. In using these priors, we are following K03 (although they
do not include the prior on the $N_* - r_c$ relationship, because
unlike us they fix $r_c = 0.2\,h^{-1}$ Mpc for all clusters). We also
include three priors not used by K03, labeled Priors~IV through
VI. The first of these is very similar to Prior~I; it requires
clusters with large velocity dispersions and core radii to be rare,
taking the power-law slope of the mass function and of the
mass-observable relation as parameters. Prior~V imposes a Fermi
function cutoff on the parameters $\sigma_c$ and $r_c$, ruling out
very small values. Finally, Prior~VI puts a Gaussian prior on the
difference between the line of sight velocity of the central galaxy
and that of the cluster, and specifies that the central galaxy ought
to be around $L_*$ or brighter.

\begin{deluxetable*}{lll}
  \tablewidth{0pt}
  \tablecaption{Priors
    \label{tab:priors}}
  \tablehead{
    \colhead{Label} &
    \colhead{Formula} &
    \colhead{Comments}
  }
  \startdata
  \multirow{1}{*}{Prior I}   & $-\ln[1+(N_*/N_*^I)^2]$ & Cluster mass function \\
  \hline
  \multirow{1}{*}{Prior II} & $+\ln(\sigma_c/\sigma_c^{II})$ & Unbiased estimate of $\sigma_c$\\
  \hline
  \multirow{2}{*}{Prior III}   & $-1/2\left[\log(\sigma_c/\sigma_c^{III})-a^{III}_0\log(N_*)\right]^2/(a^{III}_1)^2$ & Empirical $N_* - \sigma_c$ correlation \\
                             & $-1/2\left[\log(r_c/r^{III}_c)-a^{III}_2\log(N_*)\right]^2/(a^{III}_3)^2$ & Empirical $N_* - r_c$ correlation \\
  \hline
  \multirow{2}{*}{Prior IV}  & $+[\alpha_{\sigma}(1-\gamma)-1]\ln(\sigma_c/\sigma_c^{IV})$ & Mass function and $M_{cl} - \sigma_c$ relation\\
                             & $+[\alpha_{r}(1-\gamma)-1]\ln(r_c/r^{IV}_c)$ & Mass function and $M_{cl} - r_c$ relation\\
  \hline
  \multirow{2}{*}{Prior V}  & $-\ln(1+\exp[(\sigma_c^{V}-\sigma_c)/\Delta\sigma_c^{V}])$ & Fermi function enforcing $\sigma_c \gtrsim \sigma_c^{V}$ \\
                             & $-\ln(1+\exp[(r^{V}_c-r_c)/\Delta r^{V}_c])$ & Fermi function enforcing $r_c \gtrsim r^{V}_c$ \\
  \hline
  \multirow{2}{*}{Prior VI}  & $-1/2(v_{pec}/v^{VI}_{pec})^2$ & Peculiar velocity of central galaxy\\
                             & $-2.5a^{VI}\log(1+L_*/L_{cen})$ & Brightness of central galaxy
  \enddata
\end{deluxetable*}

\begin{deluxetable}{lccc}
  \tablewidth{0pt}
  \tablecaption{Prior parameter values
    \label{tab:priorparams}}
  \tablehead{
    \colhead{Label} &
    \colhead{Parameter} &
    \colhead{Iter. 1 Value} &
    \colhead{Iter. 2 Value}
  }
  \startdata
  \multirow{1}{*}{Prior I} & $N_*^I$ & $0.1$ & $0.1$ \\
  \hline
  \multirow{1}{*}{Prior II} & $\sigma_c^{II}$ & $1000~\mathrm{km}~\mathrm{s}^{-1}$ & $1000~\mathrm{km}~\mathrm{s}^{-1}$ \\
  \hline
  \multirow{6}{*}{Prior III} & $a^{III}_0$ & $0.526$ & $0.445$ \\
                           & $a^{III}_1$ & $0.0744$ & $0.153$ \\
                           & $a^{III}_2$ & $0.114$ & $0.155$ \\
                           & $a^{III}_3$ & $0.071$ & $0.037$  \\
                           & $\sigma_c^{III}$ & $254~\mathrm{km}~\mathrm{s}^{-1}$ & $250~\mathrm{km}~\mathrm{s}^{-1}$ \\
                           & $r^{III}_c$ & $0.231~\mathrm{Mpc}$ & $0.203~\mathrm{Mpc}$ \\
  \hline
  \multirow{5}{*}{Prior IV} & $\gamma$ & $1.8$ &  $1.8$\\
                            & $\alpha_{\sigma}$ & $2.0$ & $2.0$ \\
                            & $\alpha_{r}$ & $2.25$ & $2.25$ \\
                            & $\sigma_c^{IV}$ & $800~\mathrm{km}~\mathrm{s}^{-1}$ & $800~\mathrm{km}~\mathrm{s}^{-1}$ \\
                            & $r^{IV}_c$ & $0.2~\mathrm{Mpc}$ & $0.2~\mathrm{Mpc}$ \\
  \hline
  \multirow{4}{*}{Prior V} & $\sigma_c^{V}$ & $200~\mathrm{km}~\mathrm{s}^{-1}$ & $200~\mathrm{km}~\mathrm{s}^{-1}$ \\
                            & $\Delta \sigma_c^{V}$ & $10~\mathrm{km}~\mathrm{s}^{-1}$ & $10~\mathrm{km}~\mathrm{s}^{-1}$ \\
                            & $r^{V}_c$ & $0.1~\mathrm{Mpc}$ & $0.1~\mathrm{Mpc}$ \\
                            & $\Delta r^{V}_c$ & $0.01~\mathrm{Mpc}$ & $0.01~\mathrm{Mpc}$ \\
  \hline
  \multirow{2}{*}{Prior VI} & $v^{VI}_{pec}$ & $200~\mathrm{km}~\mathrm{s}^{-1}$ & $200~\mathrm{km}~\mathrm{s}^{-1}$ \\
                            & $a^{VI}$ & $1.0$ & $1.0$ 
  \enddata
\end{deluxetable}

Some of the prior parameters listed in Table~\ref{tab:priorparams} are
simply constant offsets to the likelihood, and are thus unimportant
except insofar as they may admit or exclude clusters with likelihoods
near the cutoff value. The parameters $\sigma_c^{II}$,
$\sigma_c^{IV}$, and $r_c^{IV}$ fall in this category. We set them to
nominal values, adopting K03's value for $\sigma_c^{II}$. For Priors~I
and III, we use the same parameter values as K03 for the first cluster
search iteration, making a guess for the values for the relationship
between $N_*$ and $r_c$. For Priors~IV and V we use theoretically
motivated guesses at the parameter values. In particular, Prior~IV
encodes the expectation that $dN/dM \propto M^{-\gamma}$ and that $M
\propto \sigma_c^{\alpha_\sigma} \propto
r_c^{\alpha_r}$. \citet{Tinker:2008p709} parameterize the cluster mass
function using a more complicated functional form, but we can
approximate it as a power law of slope $\gamma=-1.8$.

\begin{figure}
\includegraphics[width=\columnwidth]{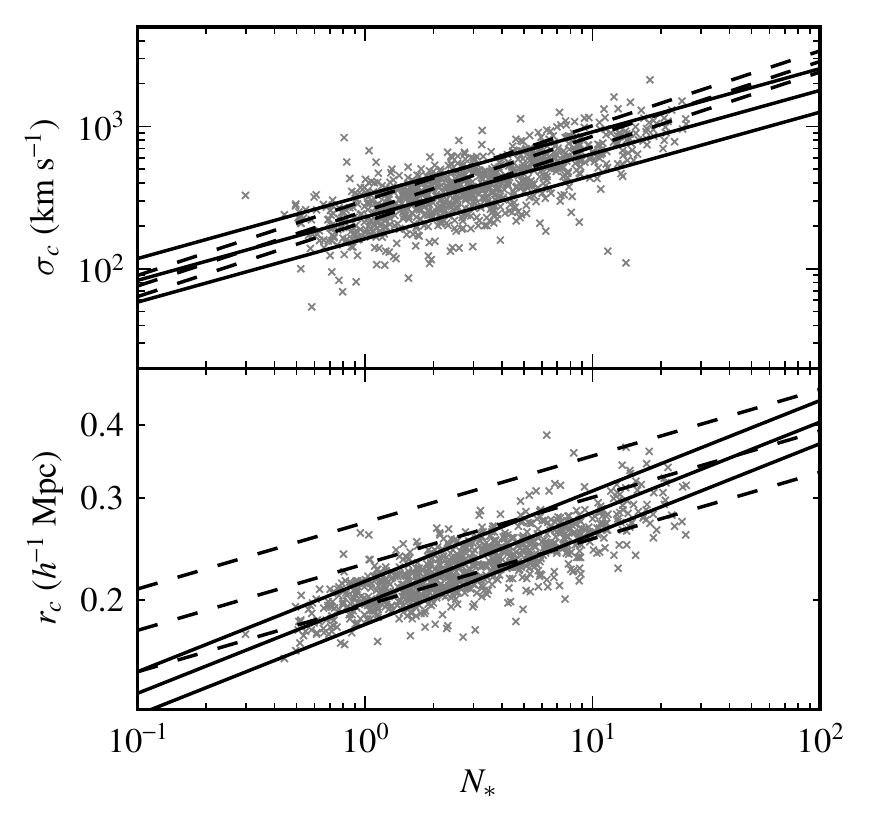}
\caption{Velocity dispersion $\sigma_c$ (first panel) and core radius
  $r_c$ (second panel) versus richness $N_*$, for selections of
  clusters from the first cluster search iteration. The dashed lines
  indicate the relationship used in Prior~III during the first search
  iteration, with its $1\sigma$ width, and the solid lines indicate
  the relationship and width used during the second iteration. See
  \protect{Section~\ref{sec:priors}}, as well as
  \protect{Tables~\ref{tab:priors} and \ref{tab:priorparams}}.}
\label{fig:priorcorr}
\end{figure}

Our third prior reflects empirical relationships between cluster
richness and other properties. We examine the correlations between
these properties in subsets of the clusters found in our first cluster
search iteration, in order to update the parameters of this prior for
the second iteration. We turn first to the $N_* - \sigma_c$
correlation. For this test we used a subset of 902 clusters with
likelihoods $\Delta \ln \mathcal{L} > 20$, expected number of visible
galaxies $N_{g666} > 3$, and measured velocity dispersions (i.e., we
excluded clusters with too few redshift measurements). The top panel
of Figure~\ref{fig:priorcorr} shows the distribution of these
clusters. We find a strong correlation, with a Pearson coefficient $r
= 0.75$. The prior used in the first iteration is plotted with its
$1\sigma$ contours, and matches the observed distribution well. We
also plot a fit to the data, with lines indicating the scatter about
the fit. The best-fit relationship is $\log(\sigma_c) =
(2.36\pm0.05)+(0.445\pm0.088)\log(N_*)$, with a scatter of $0.153$
dex. The updated parameters for Prior~III that result from this fit
are listed in Table~\ref{tab:priorparams}. Because the meaning of
$N_*$ changes between iterations, we use the transformation factor
calculated in Section~\ref{sec:derived} to express the empirical
relationships in terms of the second-iteration richness. This simply
results in a slightly different offset to the relation; these offsets
are reflected in the values of $\sigma_c^{III}$ and $r_c^{III}$
reported in Table~\ref{tab:priorparams}. When we examine the same
relationship after the second cluster finding iteration, the best-fit
relationship and the scatter are essentially unchanged.

We examine next the correlation between $N_*$ and $r_c$, using the
same subsample of clusters as before but also including 27 clusters
with unmeasured velocity dispersions. As seen in the bottom panel of
Figure~\ref{fig:priorcorr}, there is again a strong correlation
(Pearson $r = 0.77$). We plot the prior used in the first
cluster-finding iteration. The observed correlation is offset from the
prior relation, and has a scatter smaller than the width of the
prior. We find that if we adjust the prior in the second search
iteration to match this empirical relationship, a similar offset and
further reduced scatter are apparent in the second-iteration
clusters. This indicates that the galaxy distribution is not
constraining $r_c$ for individual clusters, but that it is being
determined predominantly by the priors (particularly by the
combination of Prior~III and Prior~IV, which favors smaller values of
$r_c$). This is not particularly surprising, since the detailed radial
density profile of a single cluster cannot be well constrained unless
a great many member galaxies are detected. Therefore we abandon the
strategy of determining the parameters of Prior~III using individual
$r_c$ values and turn to the stacked profiles described in
Section~\ref{sec:profile} and shown in the first panel of
Figure~\ref{fig:profile}. We fit the projected NFW profile from
Equation~\ref{eqn:gnfw} to the stacked profiles in each of the five
richness bins, varying the normalization and the scale radius
$r_c$. We restrict our fits to radii smaller than $1\,h^{-1}$ Mpc, and
ignore the innermost radial bin, which is biased high because of the
central galaxy in each cluster. We use the resulting values and
uncertainties in $r_c$ to perform a power-law fit for the relationship
between $\langle N_* \rangle$ (the average $N_*$ in each richness bin)
and $r_c$. We obtain a logarithmic slope of $0.155\pm0.068$ and
normalization (i.e., $r_c$ for $N_*=3$) of $0.235\pm0.012$. Since this
relationship is determined using the stacked profiles of actual
clusters, it is not directly affected by Prior~III; nevertheless, it
does not differ wildly from the first-iteration relationship. In our
second cluster search iteration, we update the parameters of Prior~III
to these best-fit values (see Table~\ref{tab:priorparams}). We set the
width of the prior $a_3^{III}$ to three times the uncertainty in the
normalization, to account for the additional uncertainty in the slope.

\section{Conclusions}
\label{sec:conclusions}

We follow up on the search for galaxy clusters in the 2MASS catalog
described by K03, using an iterative process to check and adjust
several of the adjustable aspects of the algorthm. The most important
component of the search process is the matched filter itself, that is
the description of the characteristics of clusters. We check the
projected density profile shape and velocity distribution of clusters,
and the luminosity function of their galaxies by stacking clusters in
bins of richness. Overall, we find that the cluster model used by K03
is mostly accurate, with radial profiles closely matching the
projected NFW model at radii less than $1\,h^{-1}$ Mpc and velocity
distributions matching the expected Gaussian distributions very
well. At large radii, out to $\sim 20\,h^{-1}$ Mpc, the observed
density profile lies above the projected NFW profile. We attribute the
excess density to correlated structure and construct a toy profile to
fit this ``two-halo'' term, but because our matched filter only
searches for galaxies within $1\,h^{-1}$ Mpc we do not bother to add
it to our matched filter. The main discrepancy between the stacked
clusters and the matched filter that was used to find them is in the
luminosity function, which we find to underestimate the fraction of
bright galaxies. After updating the matched filter with the best-fit
Schechter function, we find that the second-iteration clusters match
the filter well. The best-fit function has $M_{K*}=-23.64\pm0.04$ and
$\alpha=-1.07\pm0.03$. Though a single Schechter function fits the
data reasonably well, we find a slightly better fit when we add a
Gaussian component at the bright end, suggesting that a separate
population of BCGs is present. Based on our best guess of the BCGs in
a subset of our clusters, we estimate that the Gaussian is centered at
a $K$ magnitude of $-24.57$, with a variance of ($0.59$
mag)$^2$. Including this component causes the Schechter parameters to
change considerably; their new best-fit values are
$M_{K*}=-23.27\pm0.07$ and $\alpha=-0.89\pm0.05$. The Gaussian BCG
component peaks at a value of $(0.048\pm0.009)\phi(M_{K*})$. We do not
find that the average profiles, velocity distributions, or luminosity
functions of clusters varied with $N_{g666}$, the (distance-dependent)
number of cluster galaxies that we expect to be brighter than the
survey limit.

We also update the priors that are added to the likelihood
function. We include the three priors used by K03, including one
(labeled Prior~III) which takes into account the empirical
relationship between richness and velocity dispersion and adding one
for the analogous relationship between richness and core radius. We
also add three more priors in an effort to improve the completeness
and purity characteristics of the cluster sample. The first of these,
which we label Prior~IV, discourages clusters with large velocity
dispersions and core radii, as they are associated with (rare) massive
clusters. Another puts lower limits on the values that these variables
can take, and the last puts a prior on the central galaxies of
clusters, encouraging them to be brighter than $L_*$ and to have small
peculiar velocities. We use the clusters found in our first cluster
search iteration to tune the empirical relationships on which
Prior~III is based for the second iteration.

After adjusting the parameters of the matched filter and the priors,
we repeat our search for clusters. The richness and redshift
distributions of the resulting sample of 7624 cluster candidates with
$\Delta \ln \mathcal{L} \ge 5$ (2087 with $N_{g666} > 3$) are shown in
Figure~\ref{fig:zrichscatter}. This is a larger sample than the 5793
(1532 with $N_{g666} > 3$) found in the first iteration. When we
repeat the stacking analysis on this second catalog, we find that the
shape of the filter is well-matched to the average properties of the
clusters. We further subdivide the clusters into membership samples
with different values of $N_{g666}$ in order to test for evolution in
the cluster parameters with the number of detected cluster galaxies,
and find no strong evidence for such evolution.

We mention in closing two technical points. First, we have introduced
the general approach for studying clusters of ``catalog bootstrap
resampling''. By simultaneously resampling both the cluster catalog
and the galaxies, we can include many statistical uncertainties in a
well-defined manner. This bootstrap approach could also be applied to
the galaxy catalog during the process of finding clusters, where the
variance in the resulting cluster catalogs and properties would probe
many, though not all, of the systematic problems associated with
identifying clusters, and would yield meaningful constraints on the
purity of the catalog. The one operational issue is that repeated
galaxies should be spatially shifted away from one another in order to
avoid overly artificial density spikes, but not by so much that
density profiles are overly smoothed. This suggests scales of order
$r_c$, but tests in artificial catalogs can be used to test and
optimize this scheme. Second, while in this work we have stacked
clusters in order to ``manually'' adapt the matched filter used to
find clusters, the process could in principle be automated. For
instance, once an initial catalog is found, one could adjust the
parameters of the matched filter to maximize the overall likelihood
value based on the fixed properties of that cluster catalog. A new
catalog could then be constructed using that updated matched
filter. Note that this is still an iterative process; it is doubtful
that an attempt to simultaneously find clusters \emph{and} optimize
the matched filter would be numerically stable.

\acknowledgements

We thank D. H. Weinberg and E. Rozo for their helpful comments, and
L. Macri and J. Huchra for access to early versions of the 2MASS
Redshift Catalog. This research is supported by NASA ADP grant
NNX07AH41G.

\bibliography{ms}

\end{document}